\documentstyle[epsfig]{article}

\begin{document}

\title{Towards Mapping the Proton unintegrated Gluon Distribution
in Dijets Correlations in Real and Virtual Photoproduction at HERA} 

\author{A. Szczurek \\
{\it Institute of Nuclear Physics, PL-31-342 Cracow, Poland}\\  
{\it E-mail: szczurek@alf.ifj.edu.pl} }

\maketitle

\begin{abstract}
It is discussed how the dijet azimuthal correlations in DIS and real
photoproduction at HERA probe the unintegrated gluon distribution in
the proton. The correlation function shows a strong dependence
on kinematical variables. We discuss a possible interplay of
perturbative and nonperturbative effects.
\end{abstract}

\section{Introduction}

The jet studies are known to be a good tool to test perturbative
QCD effects.
It was pointed out already some time ago that dijet production
in DIS could be a method to study the onset of BFKL dynamics both in photo-
\cite{FR94} and electroproduction \cite{AGKM94}.
Unfortunately in practice, due to unavoidable cuts on transverse
momenta of jets, one samples rather large values of the gluon
longitudinal momentum fraction $x_g$,
where it is not completely clear what is the underlying dynamics and
in particular what unintegrated gluon distribution should be used.

This presentation is based on Ref. \cite{SNSS2001} where we have
discussed the jet production beyond the familiar
collinear approximation and have focused on how more exclusive and
more differential jet production observables probe the unintegrated
gluon distribution. Based on the unintegrated gluon distributions
found recently \cite{IN00} from the phenomenological analysis of
$\sigma_{\gamma^*p}^{tot}$ we have explored dijet azimuthal correlations.

%----------------------
\section{The formalism}
%----------------------

At the parton level the total cross section for quark-antiquark
dijet production
$ \gamma^* + p \rightarrow j_1 + j_2 + X$
can be written in a compact way as:
\begin{equation}
\sigma_{T/L}^{\gamma^* p \rightarrow j_1 j_2}(x,Q^2) =
\int d\phi
 \int_{p_{1,\perp,min}^2} dp_{1,\perp}^2
 \int_{p_{2,\perp,min}^2} dp_{2,\perp}^2
\; \;
\frac{f_g(x_g,\kappa^2)}{\kappa^4} \cdot
 {\tilde \sigma}_{T/L}(x,Q^2,\vec{p}_{1,\perp},\vec{p}_{2,\perp}) \; ,
\label{total_dijets}
\end{equation}
where $x$ and $Q^2$ are standard kinematical variables.
In the formula above $f_g(x_g,\kappa^2)$ is the unintegrated gluon
distribution and
$\vec{\kappa}$ is the transverse momentum of the exchanged gluon.
It is related to the quark/antiquark jet transverse momenta
$\vec{p}_{1,\perp}$ and $\vec{p}_{2,\perp}$ as:
\begin{equation}
\vec{p}_{2,\perp}=\vec{\kappa} - \vec{p}_{1,\perp} \, , \, 
\kappa^2 = p_{1,\perp}^2 + p_{2,\perp}^2 + 2 p_{1,\perp} p_{2,\perp} cos \phi \; .
\end{equation}
We have written explicitly lower cuts on the transverse momenta
of jets in (\ref{total_dijets}).
The indices $T$ and $L$ refer to transverse and longitudinal photons,
respectively.
The auxilliary quantities ${\tilde \sigma}_{T/L}$
introduced in (\ref{total_dijets}) are given explicitly
in \cite{SNSS2001}.
$f_g$ is evaluated at $x_g = \frac{M_t^2 + Q^2}{W^2 + Q^2}$ where
$M_t^2 = 
   \frac{p_{1,\perp}^2+m_f^2}{z}
 + \frac{p_{2,\perp}^2+m_f^2}{1-z}$
is flavour dependent. It is obvious then that at large
transverse momenta of jets one samples larger values of $x_g$ than in
the case of total cross section.

The gluon momentum $\kappa$ is responsible for the jets being not exactly
back-to-back in contrast to the conventional collinear approximation
to leading order.
In the following we limit ourselves to the region of $x_{\gamma} \sim$ 1,
where the jets are dominantly produced from the quark box on the very top
of the gluonic ladder.

The following simple two-component Ansatz was adopted in \cite{IN00}
\begin{equation}
f_g(x_g,\kappa^2) =
  {\cal F}_{soft}(\kappa^2) \frac{\kappa_s^2}{\kappa^2+\kappa_s^2}
+ {\cal F}_{hard}(x_g,\kappa^2) \frac{\kappa^2}{\kappa^2+\kappa_h^2}
\label{decomposition}
\end{equation}
for unintegrated gluon distribution.
The parameters $\kappa_s$ and $\kappa_h$ determine the scale of
the transition from the hard to soft gluon region \cite{IN00}.
The details concerning both components can be found in \cite{IN00}.

The hard perturbative component is calculated from known conventional
DGLAP parametrizations as derivative \cite{IN00}.
The results presented in this note and in \cite{SNSS2001} were
obtained based on a recent MRST98 LO parametrization \cite{MRST98}.

The two-component structure (\ref{decomposition}) of the unintegrated
gluon distribution leads to interesting consequences for the dijet azimuthal
correlations.

%----------------
\section{Results}
%----------------

Here for illustration only two examples will be discussed.
A more complete analysis can be found in \cite{SNSS2001}.

The cross section for the dijet production strongly depends on cuts
imposed on kinematical variables. In order to better demonstrate
the effect of coexistence of perturbative and nonperturbative effects
we have imposed cuts on kinematical variables in
the so-called hadronic center of mass (HCM) sytem.
In the present purposefully simplified analysis we impose the cuts on
the parton level and avoid extra cuts in the laboratory frame. 

\begin{figure}[t]
\begin{center}
\epsfig{file=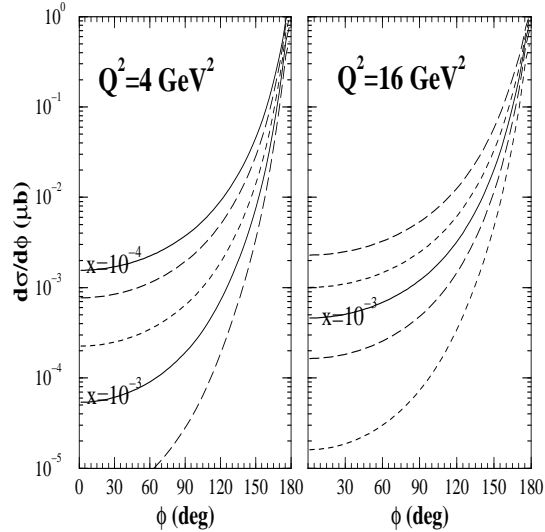, height = 6.0cm, width=6.0cm}
\end {center}
\vspace{0.5cm}
\caption{The cross section for $ \gamma^* p \rightarrow j_1 j_2 $ 
as a function of HCM azimuthal angle between jets for $Q^2$ = 4 GeV$^2$
(left panel) and $Q^2$ = 16 GeV$^2$ (right panel) for several values
of Bjorken-$x$ = 10$^{-4}$ (solid), 2 $\cdot$ 10$^{-4}$ (long-dashed),
5 $\cdot$ 10$^{-4}$ (short-dashed), 10$^{-3}$ (solid),
2 $\cdot$ 10$^{-3}$ (long-dashed),
5 $\cdot$ 10$^{-3}$ (short-dashed).
%, 10$^{-2}$ (solid).
}

\label{fig_x_dep}
\end{figure}

In Fig.\ref{fig_x_dep}, we present $d \sigma(\gamma^*p \rightarrow j_1 j_2)/d \phi$
as a function of HCM azimuthal angle between jets for two different values
of photon virtuality $Q^2$ = 4 GeV$^2$ (left panel) and
$Q^2$ = 16 GeV$^2$ (right panel) for a series of Bjorken $x$.
In this calculation, we have restricted the transverse
momenta of jets to $p_{1,\perp}^{HCM},p_{2,\perp}^{HCM} > p_{t,cut} =$ 4 GeV
and summed over light flavours $u$, $d$ and $s$.
One can observe a strong dependence of the azimuthal angle decorrelation
pattern on Bjorken $x$.
A closer inspection of both panels simultaneously leads to the conclusion
that averaging over a broad range of $Q^2$ would to a large extent destroy the
effect as it involves automatically averaging over a certain range of $x_g$,
the most crucial variable for the effect to be observed.

The experimental identification of the effects discussed here requires 
good statistics in the data sample.
In practice \cite{Maciej}, one averages over broader range of
Bjorken $x$, photon virtuality and jet transverse momenta.
Most of the effects are then washed out and
the information about the small-$x$ dynamics is to a large extent lost.

In the model in Ref.\cite{IN00} the total (real) photoproduction
cross section at energies $W <$ 100 GeV is dominated by
the soft component. Only at very high,
not yet available, energies the hard component would dominate.
At "intermediate" energy available at HERA, the two components
coexist and their fraction is a smooth function of initial
$\gamma p$ energy.
In principle, the same stays true for the dijet production and
has interesting consequences for the jet azimuthal correlations
\cite{SNSS2001}.

\begin{figure}[t]
\begin{center}
\epsfig{file=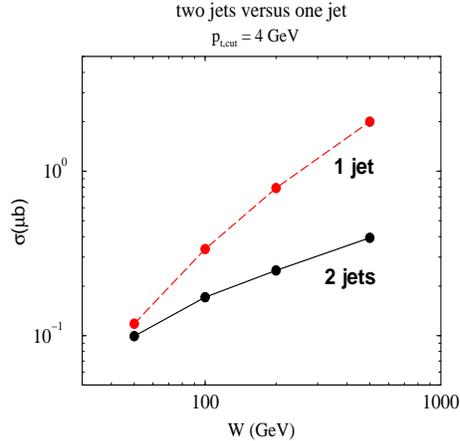, height = 6.0cm, width=6.0cm}
\end {center}

\vspace{0.1cm}
\caption{
The cross section for $\gamma p \rightarrow$ "two jet" (solid) and "one jet"
(dashed) cases as a function of $\gamma p$ CM energy. In this calculation
$p_{t,cut}$ = 4 GeV. 
}

\label{fig_1vs2}
\end{figure}

There is another interesting prediction of the two-component model.
Let us concentrate on the cases (events) with one hard
($p_{\perp} > p_{t,cut}$) jet and one soft ($p_{\perp} < p_{t,cut}$) "jet".
In this single jet event $x_{\gamma} <$ 1, because the transverse momentum
of the single quark(antiquark) jet is compensated by a transverse momentum
of a much softer gluon. 
We shall call such cases "one jet" events for simplicity.
Let us compare the rate of such "one jet" events to the previously
discussed cases of two hard jets in photoproduction.
As an example in Fig.\ref{fig_1vs2} we compare the cross section for the
two cases with the lower cut on transverse momentum $p_{t,cut}$ = 4 GeV.
Firstly, we observe that the cross section for both cases
are of similar order.
Furthermore we observe a significantly stronger rise of the cross section
for the "one jet" case than for the "two jet" case.
This is related to different $x_g$ and $\kappa$ sampled in both cases.
For example for W = 100 GeV and $p_{t,cut}$ = 4 GeV 
in the "one jet" case $<x_g> \approx$ 0.01 is substantially lower than
in the "two jet" case $<x_g> \approx$ 0.02. Both numbers are, however,
substantially larger than average $<x_g> \approx$ 0.005 sampled in the case
of total cross section. 
The effect of $\kappa$ is more complicated as averaged $\kappa$ strongly
depends on $\phi$ for the "two jet" case. The interplay of the two
effects ($x_g$ and $\kappa$) causes $f_g$
to be sampled differently in the "one jet" and "two jet" cases. This,
potentially allows the possibility of a further nontrivial test
of $f_g$. 
It would be valuable to compare the present predictions with
the predictions of standard (collinear) NLO approach.

%--------------------
\section{Conclusions}
%--------------------

Based on the recent model determination \cite{IN00} of the unintegrated
gluon distribution in the proton we have explored the impact of the soft
gluon component and the onset of the perturbative regime on
the dijet azimuthal correlations.
We have predicted  a strong dependence of the azimuthal correlation pattern
on Bjorken $x$, photon virtuality and the cut on the jet transverse momenta.
The effects in the electroproduction could be verified now at HERA, provided
a careful differential ($x$ ,$Q^2$, transverse momentum cut) studies
of the dijets are made.

It would be important to
compare the results of the model discussed here with the result
of the standard collinear approach to understand the potential of such a dijet
study to shed more light on the low-$x$ dynamics which has been studied up to
now in rather inclusive processes. Finally, we have found that the study of
the energy dependence of the "one jet" (defined in the text) cross section
would be a new test of unintegrated gluon distributions.

\section*{Acknowledgments}
I am indebted to coauthors of Ref.\cite{SNSS2001} for collaboration
on the subject of this presentation.
This work was partially supported by the German-Polish DLR exchange program,
grant number POL-028-98.

%-------------------------------------------------------------------------

\end{document}